\documentclass[useAMS,usegraphicx]{mn2e}
\usepackage{times}

\newif\ifAMStwofonts
\AMStwofontstrue

\def\ton{{Ton S180}}
\def\eqpair{{\textsc{eqpair}}}

\def\einstein{{\it Einstein}}
\def\exosat{{\it EXOSAT}}
\def\xmm{{\it XMM-Newton}}
\def\chandra{{\it Chandra}}

\def\et{{et al.\ }}

\def\heao{{\it HEAO-1} A2}
\def\asca{{\it ASCA}}
\def\sax{{\it BeppoSAX}}

\newcommand{\ls}{\mathrel{\hbox{\rlap{\hbox{\lower4pt\hbox{$\sim$}}}\hbox{$<$}}}}
\newcommand{\gs}{\mathrel{\hbox{\rlap{\hbox{\lower4pt\hbox{$\sim$}}}\hbox{$>$}}}}

\def\arcs{{\hbox{$^{\prime\prime}$}}}

\def\Msun{\hbox{$\rm ~M_{\odot}$}}

\def\H0{{\rm ~km~s^{-1}~Mpc^{-1}}}

\def\et{{et al.}}

\def\deg{^\circ}
\def\feii{{Fe~\textsc{ii}}}
\def\hb{{H$\beta$}}

\title[An \xmm\ observation of \ton]
        {An \xmm\ observation of \ton:
        Constraints on the continuum emission in ultrasoft Seyfert galaxies} 
\author[S. Vaughan \et]
        {S. Vaughan,$^{1}$ 
         Th. Boller,$^{2}$
         A. C. Fabian,$^{1}$
         D. R. Ballantyne,$^{1}$
         W. N. Brandt$^{3}$ and
         J. Tr\"umper$^2$\\
$^1$Institute of Astronomy, Madingley Road, Cambridge CB3 0HA \\
$^2$ Max-Planck-Institut f\"ur extraterrestrische Physik, Postfach 1603, 85748 Garching, Germany \\
$^3$ Department of Astronomy and Astrophysics, Pennsylvania State University, 525 Davey Lab, University Park, PA 16802, USA\\
}
\date{Accepted 2002 July 25; Submitted 2002 July 23; in original form 2002 June 12}
\pagerange{\pageref{firstpage}--\pageref{lastpage}}
\pubyear{2002}
\begin{document}
\maketitle
\label{firstpage}

\begin{abstract}
We present an \xmm\ observation of the bright, narrow-line, ultrasoft
Seyfert 1 galaxy \ton. The 0.3--10~keV X-ray spectrum is steep and
curved, showing a steep slope above 2.5~keV ($\Gamma \sim 2.3$) 
and a smooth, featureless excess of emission at lower energies.  The
spectrum can be adequately parameterised using a simple double
power-law model.  The source is strongly variable over the course of
the observation but shows only weak spectral variability, with the
fractional variability amplitude remaining approximately constant over
more than a decade in energy. The curved continuum shape
and weak spectral variability are discussed
in terms of various physical models for the soft X-ray excess
emission, including reflection off the surface of an ionised accretion
disc, inverse-Compton scattering of soft disc photons by thermal
electrons, and Comptonisation by electrons with a hybrid
thermal/non-thermal distribution.
We emphasise the possibility that the strong
soft excess may be produced by dissipation of accretion energy in the
hot, upper atmosphere of the putative accretion disc.
\end{abstract}


\section{Introduction}
\label{sect:intro}

Many Seyfert 1 galaxies and radio-quiet quasars possess X-ray spectra
that steepen below $\sim 2$~keV. This `soft excess' emission, usually
defined as the excess emission over and above the hard X-ray
power-law, was first seen in spectra obtained by \heao\ (Pravdo \et\
1981), \exosat\ (Arnaud \et\
1985; Turner \& Pounds 1989) and \einstein\ (Bechtold \et\ 1987).  Its
physical origin remains uncertain, as does its connection with the
so-called `big blue bump' emission rising through the ultraviolet,
although it is often associated with thermal emission from the putative
accretion disc (e.g. Arnaud \et\ 1985; Czerny \& Elvis 1987).

Boller, Brandt \& Fink (1996) and Laor \et\ (1997) showed that the
objects with the steepest soft X-ray spectra (implying the strongest soft
excesses) tend to have relatively narrow optical \hb\ lines; many are
classified optically as narrow-line Seyfert 1s (NLS1s; Osterbrock \&
Pogge 1985). These `ultrasoft' Seyferts often show other notable
properties such as very rapid X-ray variability (e.g. Forster \& Halpern
1996; Boller \et\ 1997; Turner \et\ 1999; Leighly 1999a; Brandt \et\
1999) and strong optical \feii\ emission (e.g. Lawrence \et\ 1997;
Vaughan \et\ 2001). The X-ray spectral form, X-ray variability and
optical broad-line properties seem interconnected.

Here we report the results of a 30~ksec \xmm\ observation of
Tonantzintla S180 ($z=0.062$) obtained as part of a guaranteed time
programme (PI: Th. Boller) to  study the  timing and spectral properties of ultrasoft
NLS1s using \xmm.  \ton\ is one of the X-ray brightest ultrasoft
Seyferts (Vaughan \et\ 1999; Leighly 1999a) and its 
luminosity is such that it is often classified as a quasar.
Previous X-ray observations
with \asca\ (Turner, George \& Nandra 1998; Vaughan \et\ 1999; Leighly
1999b; Ballantyne, Iwasawa \& Fabian 2001) and \sax\ (Comastri \et\
1998) showed a strong excess of soft emission and tentative evidence
for an  ionised iron line. The more recent high-resolution \chandra\
LETGS spectrum (Turner \et\ 2001b) showed no strong, narrow absorption
or emission features and showed the soft X-ray continuum is smooth and
featureless.

The rest of the paper is organised as follows. In the following
section the observation and basic data reduction procedures are
outlined. Section~\ref{sect:spectra} describes the X-ray spectral
fitting results. This is followed by an analysis of the variability
properties of \ton\ in section~\ref{sect:timing}.  The implications of
these results are discussed in section~\ref{sect:disco}.

\section{Observation and data reduction}
\label{sect:reduction}

\xmm\ (Jansen \et\ 2001) carries three European Photon Imaging Cameras
(EPIC), specifically two MOS (Turner \et\ 2001a) and one pn (Str\"{u}der
\et\ 2001) CCD camera. In addition there are two Reflection Grating
Spectrometers (RGS; den Herder \et\ 2001) and an Optical Monitor (OM;
Mason \et\ 2001).

\xmm\  observed \ton\ on 2000 December 14 (revolution 0186) for a
duration of $\sim$30~ksec, during which all instruments were operating
nominally.  
The MOS1 camera was in timing mode, which severely complicates the
analysis, and these data are ignored in the present paper. Both the
MOS2 and pn were operated in small window mode to reduce ``pile up''
from this relatively bright (few ct s$^{-1}$) source, and all
EPIC cameras used the medium filter.  The RGS was operated in
standard (Spectro+Q) mode.  Extraction of science products from the
Observation Data Files (ODFs) followed standard procedures using the
\xmm\ Science Analysis System v5.3 ({\sc sas}).

The raw data were processed to produce calibrated event lists and
screened to remove unwanted hot, dead or flickering pixels, and
events due to electronic noise.
Light curves extracted from these event
lists showed the
background to be stable throughout the duration of the observation.
The total amount of ``good'' exposure time selected was $28,644$~s and
$20,338$~s for MOS2 and pn, respectively, and $30,232$~s and $29,321$~s for RGS1
and RGS2. (The lower pn exposure is due to the lower ``live time'' of
the pn camera in small-window mode, $\sim$71 per cent; Str\"{u}der \et\ 2001).

Source data were extracted from circular regions of radius 35\arcs\
from the MOS2 and pn, and show no signs of pile-up.  Background events
were extracted from off-source regions.  Events corresponding to
patterns 0--12 (single--quadruple pixel events) were extracted from
MOS2 and patterns 0--4 (singles and doubles) were used for the pn
analysis, after checking for consistency with the data extracted using
only single pixel events (pattern 0). Standard redistribution matrices
({\tt m2\_r6\_all\_15.rmf} for MOS2 and {\tt epn\_sw20\_sdY9.rmf} for
the pn) were used, and ancillary response files were generated  using
{\sc arfgen v1.48.8}.  First-order RGS spectra were extracted using the
{\sc rgsproc v1.3.3} script, and appropriate response matrices were
generated using {\sc rgsrmfgen v1.44.5}. The total number of ``good''
source events extracted was $2.9 \times 10^5$ for the pn, $8.8 \times
10^{4}$ for MOS2 and $1.4 \times 10^{4}$ for each RGS.

\section{Spectral analysis}
\label{sect:spectra}

The source spectra were grouped such that each spectral bin contains
at least 20 counts, and they were fitted using the {\sc xspec v11.1} software
package (Arnaud 1996).  The quoted errors on the derived best-fitting
model parameters correspond to a 90 per cent confidence level  for one
interesting parameter (i.e., a $\Delta \chi^{2}=2.7$ criterion) unless
otherwise stated.  Values of $ H_0 = 70 $~km s$^{-1}$ Mpc$^{-1}$ and $ q_0 = 0.5 $
are assumed throughout the paper, and fit parameters are quoted for the
rest frame of the source.

\subsection{The 2.5--10~keV spectral form}
\label{sect:hard_spec}

\begin{figure}
\centering
\includegraphics[width=5.3 cm, angle=270]{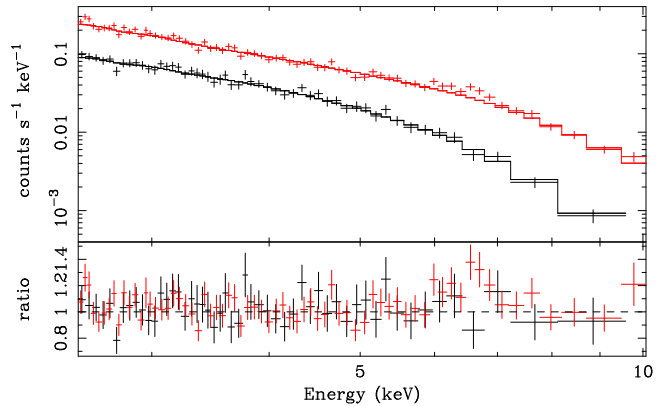}
\caption
{EPIC pn (upper data) and MOS2 spectra in the 2.5--10 keV range compared
to a simple power-law model (the energies have been shifted into the
source frame). The upper panel shows the count spectra
(crosses) and the power-law model folded through the detector
responses (histograms).  The lower panel shows the fit residuals.}
\label{fig:hard_spec}
\end{figure}

The two EPIC spectra were first examined separately to test for
discrepancies. The data were fitted over 2.5--10~keV using a model
comprised of a power-law modified by Galactic absorption (with fixed
column density $N_{H}=1.55 \times 10^{20}$~cm$^{-2}$; Dickey \&
Lockman 1990). This provided a reasonable fit to both spectra
($\chi^{2} = 115.0 / 153$~degrees of freedom and $\chi^{2} = 339.8 /
336 ~dof$ for the MOS2 and pn data, respectively) with consistent
normalisations and photon indices ($\Gamma = 2.29 \pm 0.10$ for MOS2
and $\Gamma = 2.25 \pm 0.06$ for pn).

Given this good agreement between the two spectra over the
2.5--10~keV energy range, the data were fitted simultaneously in the
rest of the spectral analysis, but in each fit the residuals from the
two spectra were examined separately to check for inconsistencies.
The relative normalisations were left free to allow for slight
differences in absolute flux calibration  between the two
detectors.  Figure~\ref{fig:hard_spec} shows the two EPIC spectra
compared to a simple power-law fitted over 2.5--10~keV. As expected from
the fits to the individual spectra, this model ($\Gamma=2.26_{-0.12}^{+0.05}$)
is in excellent agreement with the data ($\chi^{2}=455.1/490~dof$). 

\subsubsection{Iron K$\alpha$ emission}
\label{sect:iron_line}

The pn spectrum shows positive residuals around $\sim 7$~keV which could
indicate spectral features from iron K.  To test for this an
intrinsically narrow ($\sigma = 1$~eV) Gaussian emission line was
added to the model. With the narrow line energy fixed at 6.4~keV the
fit was not significantly improved; the 90 per cent upper limit of the
equivalent width of a narrow, neutral iron line is $EW<60$~eV. 
Allowing the line energy to be free provided an improvement to the fit
($\chi^{2}=447.3/488~dof$) but the best-fitting line energy of
$E=7.01\pm0.06$~keV is marginally too high for K$\alpha$ emission from
even Hydrogen-like iron.  The equivalent width of this line is
$EW=72\pm 43$~eV.  According to an $F$-test, 
upon adding two parameters (line flux and energy) the fit is
significantly improved at 98.5 per cent confidence.. 
Allowing the line width to be free improved the fit
further ($\chi^{2}=441.5/487~dof$) with a best-fitting width of
$\sigma = 0.53_{-0.33}^{+0.50}$~keV and an energy of
$E=7.01\pm0.31$~keV, which is consistent with K$\alpha$ emission
from H-like iron. An $F$-test suggests the fit is improved, at 98.8
per cent confidence, upon allowing the line width to be free.
This broad line is strong, with an equivalent width
of $EW=282_{-163}^{+245}$~eV. 
The high energy derived from the line
might instead be indicating the presence of iron K absorption
(at $E>7.1$~keV), but this seems unlikely.
Replacing the broad Gaussian with an absorption
edge gave a worse fit ($\chi^{2}=451.8/488~dof$),  with edge parameters 
$E=8.33_{-0.45}^{+0.65}$~keV and $\tau=0.15\pm0.14$.

Using a {\sc diskline} line profile (Fabian \et\ 1989) to model the residuals
around $7$~keV provided  a comparable fit to the broad Gaussian
($\chi^{2}=444.2/487$). An $F$-test suggests this improvement to the
fit, compared to the simple power-law model, is significant at $99$
per cent confidence. The best-fitting parameters of the line were the
following: rest-frame energy $E = 6.48_{-0.08}^{+0.42}$~keV,
$EW=523_{-160}^{+273}$~eV and inclination angle
$i=65_{-9}^{+21}$~deg. Allowing the other free parameters (inner
radius $r_{in}$, outer radius $r_{out}$, and disc emissivity index
$q$) to be free did not improve the fit significantly, and thus
these parameters were kept fixed at $r_{in}=6r_g$, $r_{out}=1000r_g$
and $q=3$. As can be seen from figure~\ref{fig:hard_spec}, the
excess at $\sim 7$~keV in the pn data is not obvious in the MOS2 data;
however, the two spectra are formally consistent with one another.  When
the normalisation of the {\sc diskline} was allowed to be free between
the two detectors, the best fitting normalisations were consistent,
although the normalisation form the MOS2 spectrum is poorly
constrained and is consistent with zero.

The above results are suggestive of ``reflection'' off a (possibly
ionised) accretion  disc. To provide a more physically realistic
description of emission from an ionised disc, the model of Ross \&
Fabian (1993) (see also Ballantyne \et\ 2001) was fitted to the data
(assuming solar abundances). The disc reflection spectrum was blurred
using the {\sc laor} kernel (Laor 1991) to simulate Doppler and
gravitational effects around a black hole. The best fitting ionised
disc model gives a good fit ($\chi^{2}=445.0/487~dof$) with
$\Gamma=2.43\pm0.13$, $R=1.5_{-0.9}^{+1.1}$ and fairly low  ionisation
($\log(\xi)<1.9$). The {\sc laor} profile parameters were only poorly
constrained and were kept fixed (as above).

We conclude that the presence of a broad, ionised iron
line (as previously suggested by Comastri \et\ 1998 and Turner \et\
1998) is consistent with the \xmm\ data, but the line parameters are 
only poorly constrained.

\subsection{The broad-band X-ray spectrum}
\label{sect:full_spec}

Having parameterised the spectrum above 2.5~keV the analysis was
extended to cover the full useful spectral bandpass of \xmm\ by
including the EPIC data down to 0.3~keV and the RGS data. However,
there are significant differences between the MOS2 and pn spectra
below 1~keV; the difference is most pronounced around 0.56~keV. 
There is a similar
feature present in the calibration observation of Mrk 421,
which is almost certainly related to the
instrumental O K-edge at 0.537~keV and seems to be most pronounced in
data taken in pn small-window mode (as here).  The residuals in the
MOS camera are at a much lower level ($\ls 5$ per cent) in this spectral
region.  For these reasons a conservative restriction was applied to
the pn data, the spectrum was fitted between 1.2--10~keV only, while
the MOS2 data were used over the 0.3--10~keV band.

\begin{figure}
\centering
\includegraphics[width=8.0 cm, angle=0]{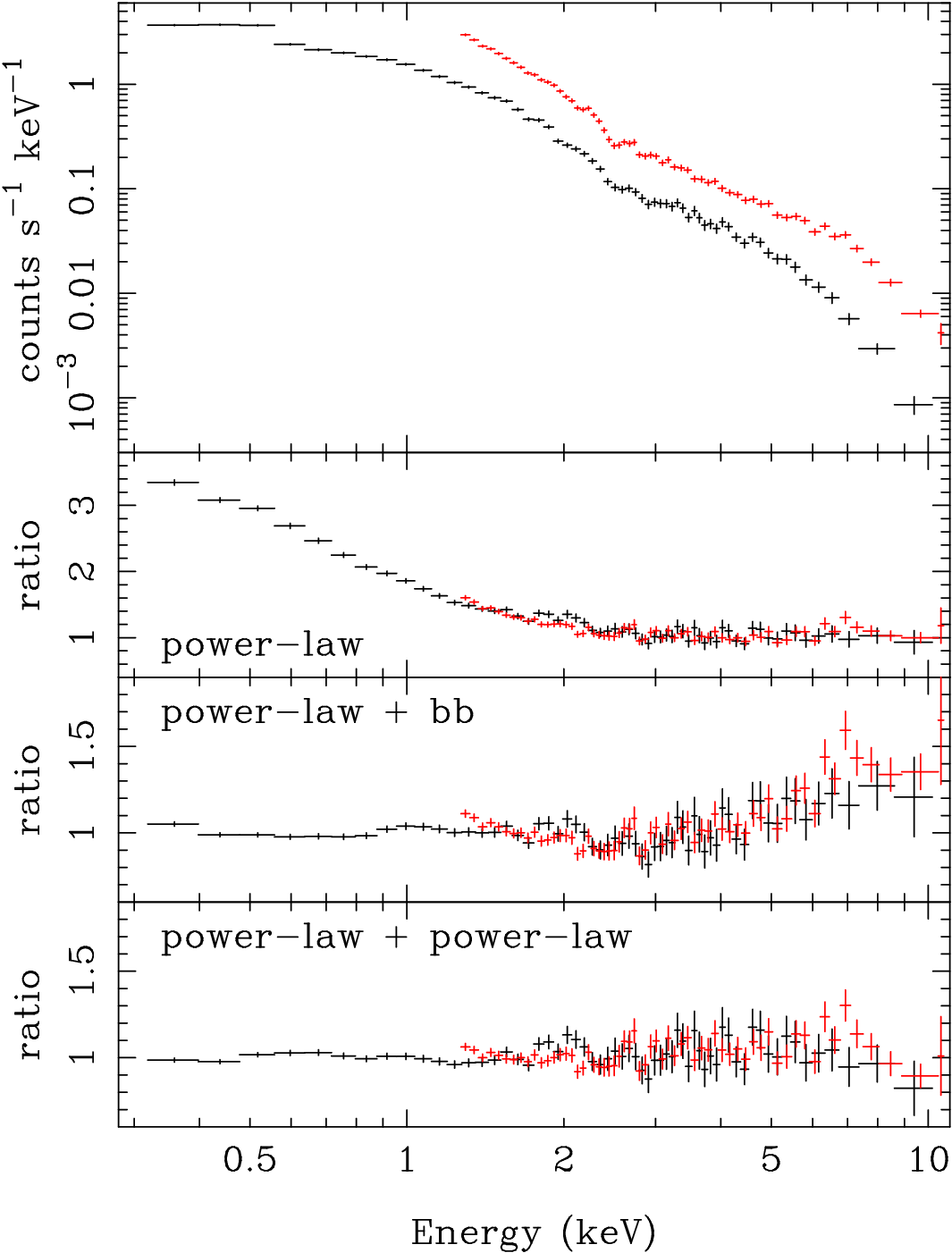}
\caption
{EPIC pn (upper data) and MOS2 spectra. The upper panel shows the
count spectra, and the lower panels show the ratio of data to three
different models.  The second panel shows the results of fitting a
single power-law model only to the data above 2.5~keV (as in
figure~\ref{fig:hard_spec}) and extrapolating  to lower energies,
revealing the curved continuum. The bottom two panels show the results
of fitting with a power-law plus blackbody or a two power-law
model. (The data were rebinned for display purposes.)}
\label{fig:full_spec}
\end{figure}

An extrapolation of the simple 2.5--10~keV power-law model down to
0.3~keV (Figure~\ref{fig:full_spec}) shows a smooth upturn in the
EPIC spectrum: the soft excess. There does not seem to be an obvious
transition from hard power-law to soft excess in the spectrum, on the
contrary, the spectrum is gently curving up to at least 2~keV. 

The average observed flux in the 0.3--10~keV band was $2.2 \times
10^{-11}$~erg s$^{-1}$  cm$^{-2}$ and the corresponding (unabsorbed)
luminosity was $\sim 2.5 \times 10^{44}$~erg s$^{-1}$. The bolometric
luminosity is most likely $\gs 10^{45}$~erg  s$^{-1}$ (this is a
conservative estimate as \ton\ probably harbours a strong EUV excess;
Turner \et\ 2002). Assuming \ton\ is radiating at the  Eddington limit
(or lower) then it requires a black hole mass $M \gs 10^{7}\Msun$ (see
also Turner \et\ 2001b).

\subsubsection{RGS spectrum}
\label{sect:rgs_spec}

\begin{figure}
\centering
\includegraphics[width=6.5 cm, angle=270]{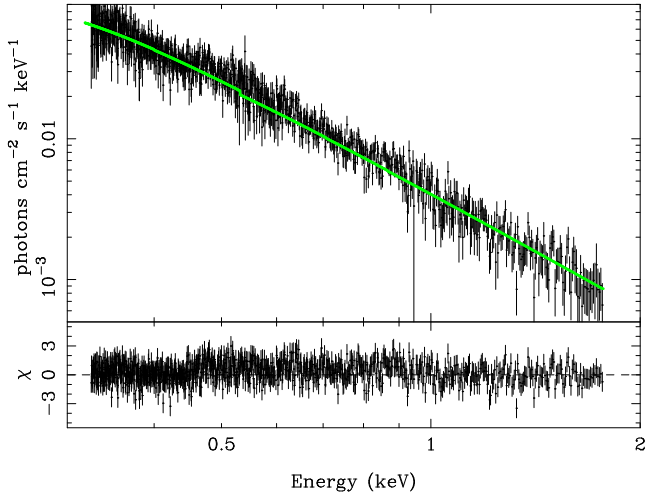}
\caption{Combined first order RGS spectrum (in ``fluxed'' units) and
best-fitting absorbed power-law model (solid line). The lower panel
shows the fit residuals.
}
\label{fig:rgs_spec}
\end{figure}

The 0.35--1.5~keV RGS data confirm that the soft excess emission in
\ton\ is a smooth continuum, as seen by Turner \et\ (2001b) from
\chandra\ LETGS data.  Figure~\ref{fig:rgs_spec} shows the 
background subtracted RGS
spectrum of \ton\ in flux units (RGS1 and RGS2 combined). 
An absorbed power-law provides a good
fit to the data ($\chi^{2}=1315.2/1256 ~dof$) with the absorbing
column fixed at the Galactic value and a power-law slope of
$\Gamma=2.77\pm0.03$. There are no significant sharp residuals in the
RGS spectrum, the only obvious deviation from the power-law model is
around 0.5--0.6~keV and is probably associated with the O K-edge in
the detector response.  The equivalent widths of any narrow
absorption features are restricted to $EW<1.0$~eV (90 per cent upper limit)
in the 0.35--0.70~keV energy range, $EW<1.7$~eV in the 0.7--1.0~keV
range and $EW<10$~eV in the 1.0--1.5~keV range.  Thus the soft excess
in \ton\ is a smooth continuum resembling a power-law.  
In the following sections the
EPIC spectra are used to test various physical models for the soft
excess emission.

\subsubsection{Phenomenological model}
\label{sect:phenom}

Given that the RGS spectrum can be fitted with a steep power law, the
broad-band EPIC spectrum was first compared to a simple double power-law
model, similar to that used by Comastri \et\ (1998) in their analysis
of a \sax\ observation of \ton. Fitting with a broken power-law model gave
$\chi^{2}=919.1/898~dof$ with slopes $\Gamma_{1}=2.98\pm0.02$,
$\Gamma_{2}=2.22\pm0.10$ and a break energy $E_{\rm
br}=2.29\pm0.04$~keV. A slightly better fit was obtained using two
separate power-laws ($\chi^{2}=910.9/898~dof$) with slopes
$\Gamma_{1}=3.11\pm0.04$ and $\Gamma_{2}=1.49\pm0.15$ (see
figure~\ref{fig:full_spec}). In this model the softer of the two power-laws
dominates the spectrum up to  $\sim 4$~keV. The harder power-law is
considerably flatter than in the broken power-law parameterisation,
due to the addition of the steeper power-law over the whole band,
and is perhaps unrealistically flat.
These simple fits to the EPIC data, and the RGS
results, show that the soft excess has the form of a steep power-law.

\begin{figure}
\centering
\includegraphics[width=6.5 cm, angle=270]{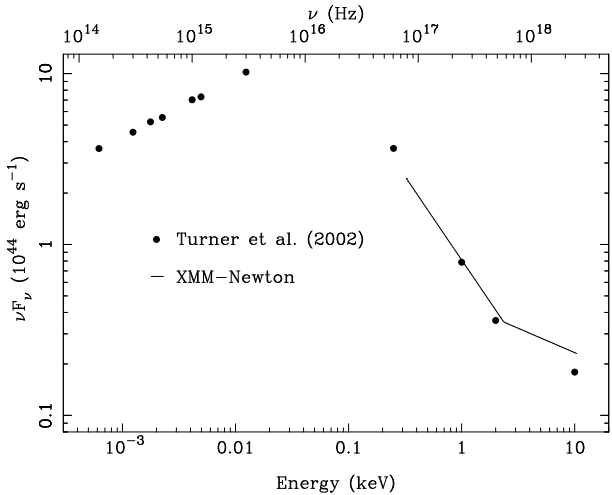}
\caption{
Spectral energy distribution of \ton, corrected for Galactic
absorption. 
The solid curve shows the \xmm\ spectrum 
(using the broken power-law parameterisation discussed
in section~\ref{sect:phenom}).
The dots show the multiwavelength data from Turner \et\ (2002)
corrected to match the  slightly different
cosmology used here. 
}
\label{fig:sed}
\end{figure}

Figure~\ref{fig:sed} places
the \xmm\ spectrum in the context of the 
multiwavelength spectral energy distribution (SED) compiled by Turner
\et\ (2002). The multiwavelength data were taken approximately one
year before the \xmm\ observation, but it is clear that the X-ray
spectral form did not change drastically between observations.

\subsubsection{Ionised reflection}
\label{sect:refl_spec}

The possible presence of iron K$\alpha$ emission
(section~\ref{sect:iron_line}) may indicate significant reflection off
the surface of the accretion disc. The reflection spectrum can be
strong below $\sim 1$~keV and could contribute significantly to the
soft excess.  To test whether the soft excess in \ton\ could
be produced by reprocessing of primary X-rays in the disc the
reflection model discussed in section~\ref{sect:iron_line} was fitted
to the 0.3--10~keV spectrum.  The best fitting model obtained for the
data above 2.5~keV gave a very poor fit to the broad-band data
($\chi^{2}=8351.1/897~dof$). The ionisation state of the reflector
means it produces strong O~{\sc vii} recombination emission, not
obvious in the data, while the total reflected flux below 2~keV falls
far short of the observed soft excess.

Refitting this model to the full-band EPIC data still provided a rather poor
fit to the data ($\chi^{2}=1030.4/896$).  The best fitting parameters
were $\Gamma=2.58\pm0.03$, $R=1.12_{-0.12}^{+0.08}$, $\log(\xi)=
3.71\pm0.06$ and parameters of the {\sc laor} kernel were $r_{\rm in}
< 2.5 r_{\rm g}$ and $i=46\pm8 \deg$ ($r_{\rm out}$ was kept fixed at
$400r_{\rm g}$).  In this model the disc surface is
highly ionised and produces a smooth excess of emission in the
soft band, partly from bremsstrahlung in the irradiated skin of the
disc, with relatively weak spectral features. However, this model
leaves significant residuals in the fit above 7~keV, where it predicts
a strong, ionised iron edge. 
Thus the low-energy data seemed to require a highly ionised
reflector, while the higher-energy spectrum suggested a reflector with
much lower ionisation. 


Allowing for two reflection components, with different
ionisation parameters, improved the fit ($\chi^{2}=964.7/894~dof$).
In this fit the continuum slope was $\Gamma=2.61\pm0.02$ incident 
on reflectors with ionisation parameters $\log(\xi_{1})=4.4\pm0.3$ and 
$\log(\xi_{2})=1.8\pm0.1$, the {\sc laor} profile parameters were
$r_{\rm in} < 1.9 r_{\rm g}$ and $i=41\pm6 \deg$. In this model the
highly ionised reflector had relative strength $R_{1}\approx 0.5$ while the
colder reflector was much stronger, with $R_{2} \approx 2.4$. 
Both reflection components contributed to the soft excess
emission, but the strong colder reflector improved the fit to the
hard X-ray data, although significant residuals still remained above 7~keV.
The small inner radius of the {\sc laor} kernel was required in
order to smooth sufficiently the recombination features present
in the reflection spectra.

\subsubsection{Purely thermal soft excess}
\label{sect:thermal_spec}

A power-law plus single blackbody component, to model the soft excess,
provided a poor fit to the data ($\chi^2=1123.6/898~dof$, 
see figure~\ref{fig:full_spec}). Including a second blackbody component
significantly improved the fit ($\chi^2=912.2/896~dof$), while a
power-law plus three blackbodies provided an excellent fit
($\chi^2=887.8/894~dof$) with the following parameters:
$\Gamma=2.24\pm0.02$, $kT_{1}=33_{-9}^{+27}$~eV, $kT_{2}=103\pm5$~eV,
$kT_{3}=248\pm15$~eV. This multiple blackbody parameterisation is
similar to that used to characterise the broad-band X-ray spectra of  other
ultrasoft Seyferts, e.g., RE~J1034+396 (Pounds, Done \& Osborne,
1995), PKS 0558--504 (O'Brien \et\ 2001) and 1H 0419--577 (Page \et\ 2002). 

However, while this multiple blackbody model did provide an adequate
fit to the data, the temperatures and luminosities of these components
are at odds with those expected for a simple accretion disc.   A
standard accretion disc around a $10^{7}\Msun$ black hole should
have a peak temperature $\ls 40$~eV, much lower than the hottest
temperature in the multiple blackbody model. The observed
temperatures and luminosities predict sizes of the blackbody emitting
regions of $r_{\rm BB} \sim 3 \times 10^{10}$~cm for the hottest and
$r_{\rm BB} \sim 10^{13}$~cm for the coolest, which are also difficult
to reconcile with a simple accretion disc around a $10^7 \Msun$
black hole ($r_{\rm g} \sim 1.5\times 10^{12}$~cm).

Fitting the data with a more realistic thermal accretion disc
spectrum, using the {\sc diskpn} code (Gierli\'{n}ski \et\ 1999),
provided a poor fit to the data ($\chi^{2} = 1073.7/898~dof$), because
the model spectrum is not broad enough,  and again it gave an
unreasonably high temperature ($kT=131\pm5$~eV).  
The strong, rapid variability of the soft excess
(section~\ref{sect:timing}) also seems incompatible with the expected
timescales in a simple accretion disc (see also Boller \et\ 1997).
These results would
seem to reject an origin for the soft excess in terms of (un-modified)
thermal emission from an accretion disc.  But, as discussed by
e.g. Bechtold \et\ (1987), Czerny \& Elvis (1987) and Ross, Fabian \&
Mineshige (1992), 
inverse-Compton scattering in the upper layers of the disc atmosphere
should significantly affect the observed high-energy tail of the disc
emission. This possibility is discussed below.

\subsubsection{Thermal Comptonisation}
\label{sect:compt_spec}

Inverse-Compton scattering of soft photons in an optically thick
medium (i.e., $\tau > 1$; where the diffusion approximation is valid)
produces an approximately power-law spectrum at low energies.  Thus,
as the soft excess in \ton\ resembles a steep power-law, a Comptonised
blackbody component should provide a reasonable fit.  The energy index of the
Comptonised spectrum is determined by the Compton $y$-parameter ($y
\approx 4 \Theta \tau^{2}$, where $\Theta = kT/m_{\rm e}c^{2}$ is the
dimensionless electron temperature) according to $\alpha \approx
\sqrt{9/4+4/y}-3/2$ (e.g., section 7.7 of Rybicki \& Lightman, 1979).
This means that it is not possible to solve for both $\tau$ and $kT$
from knowledge of the spectral slope alone.  Using the double
power-law parameterisation of \ton\ (section~\ref{sect:phenom}) gave a
slope of $\alpha = 2.1$ for the soft excess, which  corresponds to $y
\approx 0.37$ and it is possible to obtain this value, and hence the
observed spectral slope, for a range of
$\tau$ and $kT$ values.  A similar situation occurs when considering
the optically thin regime, in which case the spectral slope is
determined by $\alpha \approx -
\ln{\tau}/\ln{(1+4\Theta+16\Theta^2)}$.  See Zdziarski (1985) and
references therein for more details.

The {\tt CompTT} code (Titarchuk 1994) was used to model
Comptonisation of soft photons in a thermal plasma. A
power-law plus Comptonisation component gave a good fit
to the data ($\chi^{2}=903.7/896~dof$) with $\Gamma=1.64\pm0.18$ and the
soft excess modelled by Comptonisation of a thermal spectrum from a
$kT_{\rm bb}=60\pm7$~eV  source.
However, for the reasons mentioned above,
the temperature and optical depth of the Comptonising plasma are
strongly covariant parameters, and thus cannot be constrained simultaneously.

Fitting the spectrum with just two Comptonisation components, similar
to the model used by O'Brien \et\ (2001) and Page \et\ (2002), gave a
good fit ($\chi^{2}=891.9/895~dof$). The seed photon 
temperature was $kT_{\rm bb}=56_{-11}^{+6}$~eV, but the
parameters of the two plasmas were poorly constrained due to the degeneracy mentioned
above. (Similar fits were obtained using the \eqpair\ code, discussed
below, to model two purely thermal plasmas.)
It seems reasonable to
associate the source of the soft seed photons with the
thermal emission from the inner regions of an optically thick
accretion disc. The  existence of two discrete Comptonising plasmas
is not so easy to explain, especially given that they varied almost
identically on short timescales (section \ref{sect:timing}).

\subsubsection{Non-thermal Comptonisation}
\label{sect:compt2_spec}

The double Comptonisation model described above has the two
spectrally identified continua (soft and hard power-laws) originating
in two distinct thermal Comptonising plasmas.  An alternative is that
the whole spectrum is produced by a single plasma with a hybrid
thermal/non-thermal electron distribution (see Coppi 1999 for a
discussion of the physics of such plasmas). Gierli\'{n}ski \et\ (1999)
modelled the spectrum of the Galactic Black Hole Candidate (GBHC)
Cygnus X-1 in its high/soft state in terms of a hybrid plasma.  In
this model the Cyg X-1 spectrum is comprised of a blackbody from the
accretion disc, plus a component due to inverse-Compton scattering of
the disc photons in a corona containing both thermal and non-thermal
electrons.  The  non-thermal tail to the electron distribution
produces the steep power-law tail of the X-ray spectrum dominating
above 10~keV (and observed up to at least $\sim 800$~keV) and the
thermal electrons produce a ``soft excess'' at a few keV between
the non-thermal tail and the accretion disc spectrum.   The definite
detection of a non-thermal Comptonisation component requires
high-energy data, which are not available here, but, as first noted
by Pounds \et\ (1995), the X-ray spectra of ultrasoft Seyferts
do resemble that of Cyg X-1 in its high/soft state. Thus it seems
reasonable to test whether a hybrid thermal/non-thermal plasma is a
viable model for the X-ray continuum of \ton.

The \eqpair\ code (Coppi 1992; Gierli\'{n}ski \et\ 1999) was used to compute
the spectrum formed by Comptonisation of a blackbody photons in a
plasma with an hybrid electron distribution. Non-thermal
(relativistic) electrons are injected (by some unspecified process) as
a power-law in Lorentz factor, $\gamma$, into a background thermal
plasma, the rate of injection is given by: $\dot{N}_{\rm inj}(\gamma)
\propto \gamma^{-\Gamma_{\rm inj}}$ (between $\gamma_{\rm min}$ and
$\gamma_{\rm max}$).   The total  optical depth (including that
produced by e$^{\pm}$ pairs) and the equilibrium temperature of the
thermal component (after accounting for Coulomb heating by non-thermal
e$^{\pm}$ and Compton heating/cooling) are computed self-consistently
once the initial parameters have been specified.  For a detailed
discussion of the processes accounted for by this code see
Gierli\'{n}ski \et\ (1999) and Coppi (1999).  The important parameters
of the model are the following:  
the temperature of the input thermal disc spectrum $kT_{\rm bb}$; 
the soft compactness $l_{\rm s}=L_{\rm s} \sigma_{\rm T}/r m_{\rm e}c^3$,  
 which corresponds to the power supplied by soft seed photons
 ($L_{\rm s}$ is the total seed photon luminosity and $r$ is the
 radius of the emission region);  
the similarly defined thermal compactness $l_{\rm th}$, 
 corresponding to the heating supplied to thermal electrons;  
the non-thermal compactness $l_{\rm nth}$, 
 which gives the rate of injection of non-thermal electrons; 
the index of the non-thermal electron injection spectrum $\Gamma_{\rm inj}$;  
and the  optical depth $\tau_{\rm p}$ of the background electron-proton plasma.

With so many parameters (some of which were covariant in the fit)
the $\chi^2$-space contained many local minima. In order to reduce the
complexity of the fit some parameters were fixed at physically reasonable
values (these are discussed further below). The details of the fitted
model parameters will obviously depend on the chosen values for those
parameters kept fixed. Thus, while the following fits demonstrate that the 
\eqpair\ model can reproduce the shape of the \xmm\ spectrum, the 
exact values of the fitted model parameters should be treated with some caution.

The total compactness $l=l_{\rm s}+l_{\rm h}$ (where $l_{\rm h}=l_{\rm
th}+l_{\rm nth}$ is the total hard compactness) is not constrained by
the \xmm\ spectra. If the spectrum extends 
out to $\gs 500$~keV a high compactness will lead to
efficient pair production, but the strongest observable consequence of
the presence of pairs is an annihilation line at 511~keV.  In the
absence of this information the soft compactness was therefore kept
fixed at $l_{\rm s}=30$ (the lower limit obtained in
section~\ref{sect:timing}) and the ratios $l_{\rm h}/l_{\rm s}$ and
$l_{\rm nth}/l_{\rm h}$ were used as free parameters in the
fitting. Assuming the corona covers the inner disc, the former is
related to the ratio of power dissipated in the corona compared to the
disc, and the latter gives the fraction of power deposited in
the corona that goes into the non-thermal component.  The radius of
the emitting region was kept fixed at $r=10^{14}$~cm, this is $\ls
70 r_{\rm g}$ for a $\gs 10^7 \Msun$ black hole, and at the above
assumed value of $l_{\rm s}$ this gives a reasonable soft X-ray
luminosity. The temperature of the seed blackbody was fixed at
$kT_{\rm bb}=50$~eV and the minimum and maximum Lorentz factors for the
non-thermal injection spectrum were $\gamma_{\rm min}=1.3$ and
$\gamma_{\rm max}=1000$ (the resulting model spectrum is largely
insensitive to the chosen value of $\gamma_{\rm max}$ when
$\Gamma_{\rm inj}>2$).

The \eqpair\ model gave a good fit
($\chi^{2}=892.9/897~dof$) with the following parameters: 
$l_{\rm h}/l_{\rm s} = 0.74\pm0.03$, 
$l_{\rm nth}/l_{\rm h}=0.78\pm0.02$, 
$\tau_{\rm p}=7.0_{-1.8}^{+0.8}$ and 
$\Gamma_{\rm inj}=2.84_{-0.15}^{+0.26}$. 
Thus there is marginally more power supplied by the seed photons than the
coronal electrons, and most of the electron heating is in the form of
non-thermal electrons which are injected with a steep spectrum. 
The soft excess
is produced by Compton scattering of the accretion disc spectrum by
thermalised electrons, and the emission above a few keV is
produced by the non-thermal component. 
These results
are similar to those obtained by Gierli\'{n}ski \et\ (1999) for the
high/soft state spectrum of Cyg X-1;
the significant difference is
that in this case the optical depth of the background plasma is much
higher than in Cyg X-1. If the optical depth is kept fixed at $\tau_{\rm
p}=1$ a slightly poorer (but still acceptable) fit is obtained
($\chi^{2}=910.0/898~dof$). Again the soft excess is produced by
Compton scattering of seed photons by thermal electrons, and the
harder X-ray emission is produced by the non-thermal electrons.

\section{Timing analysis}
\label{sect:timing}

This section describes timing analyses of the EPIC data.  Results are
presented only for the pn data, which have the highest
signal-to-noise, but the results were checked against the MOS2 data and
the two were found to be entirely consistent.  The ($1 \sigma$) errors
on the light curves were calculated using  counting statistics.

\begin{figure}
\centering
\includegraphics[width=6.5 cm, angle=270]{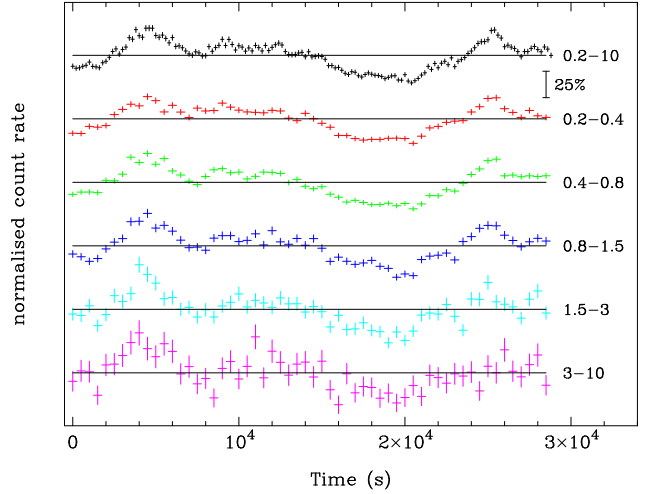}
\caption{EPIC pn light curves 
normalised to unit count rate.
The topmost panel shows the full-band 0.2--10.0~keV light curve (in
200~s bins) and
the lower panels show the light curves from various energy
bands (in 500~s bins). The vertical bar indicates the (linear) $y$-axis scale.
The mean count rates for the six light curves (from top to bottom) are as
follows: 9.26, 3.26, 3.53, 1.69, 0.56 and 0.22 ct s$^{-1}$.
}
\label{fig:curves}
\end{figure}

\begin{figure}
\centering
\includegraphics[width=6.5 cm, angle=270]{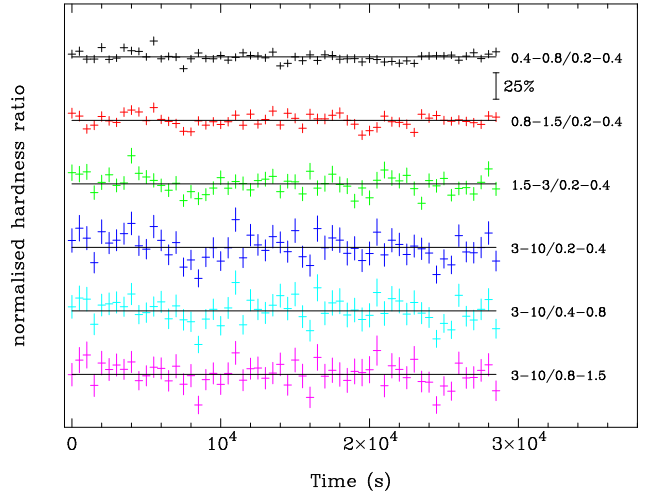}
\caption{Hardness ratios for various
EPIC pn light curves using 500~s time bins.
}
\label{fig:ratios}
\end{figure}

The topmost panel of figure~\ref{fig:curves} shows the broad-band
(0.2--10~keV) EPIC pn light curve of \ton\ in 200~s time bins. The
light curve is perfectly evenly sampled,  uninterrupted and shows a
peak-to-peak variation of $\sim 50$ per cent  over the course of the
observation. The fastest continuous rise in the light curve occurred at
around $2.4 \times 10^{4}$s, where the flux increased by $\sim 30$ per cent
in 2000~s, corresponding to an increase in luminosity of $\Delta L
/ \Delta t \approx 3.8 \times 10^{40}$~erg s$^{-2}$ in the 0.3--10~keV
band.  This was used to estimate the radiative efficiency $\eta$ of
the source, assuming photon diffusion through a spherical mass of
accreting matter, following Fabian (1979): $ \Delta L / \Delta t \ls
\eta \cdot 2.1 \times 10^{42}$~erg s$^{-2}$. (But see also Brandt \et\
1999, for a discussion of the assumptions used in this calculation.)
This gave a limit of $\eta \gs 1.8$ per cent.
The rapid variability was also used to place a limit on the
compactness parameter, following Done \& Fabian (1989). Assuming
that $r=c \Delta t$ gives $\Delta L / \Delta t \ls l \cdot
10^{39}$~erg s$^{-2}$. For \ton\ this gave $l \gs 34$.

Figure~\ref{fig:curves} also shows the pn light curves extracted from
five different energy bands in 500~s bins. The light curves all show
similar trends, albeit with relatively poor signal-to-noise in the
harder bands.  The hardness ratios of these light curves were examined
as a first test of spectral variability.  The ten hardness ratios (six
of which are shown in figure~\ref{fig:ratios}) from the five light
curves were each compared to a constant using a $\chi^{2}$ test.  Four
of the ten hardness ratios examined were inconsistent with a constant
hypothesis at $>99$ per cent confidence. These ratios, namely
$(0.4-0.8)/(0.2-0.4)$, $(0.8-1.5)/(0.2-0.4)$, $(1.5-3)/(0.2-0.4)$ and
$(1.5-3)/(0.4-0.8)$, are also among the highest signal-to-noise
ratios.  Using the Spearman rank-order  correlation coefficient and
the  Kendall $\tau$--statistic (see e.g.,  Press \et\ 1992) to test for
correlations, none of
the hardness ratios was correlated with the broad-band count rate at
greater than 90 per cent confidence.

\begin{figure}
\centering
\includegraphics[width=6.5 cm, angle=270]{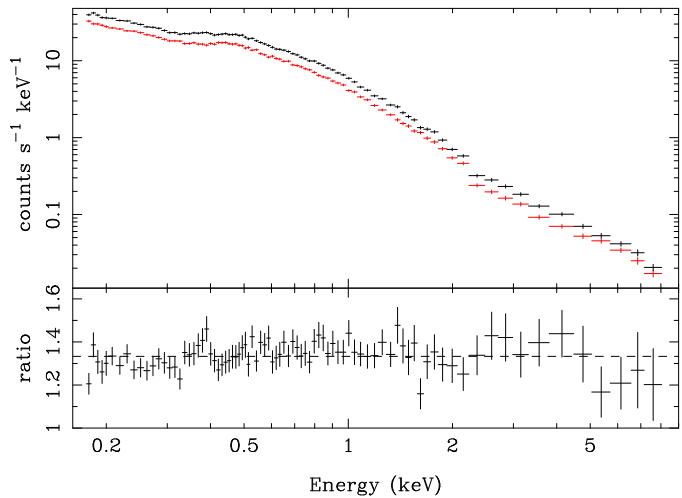}
\caption{
Comparison of high and low flux spectra. The top panel shows the two
count spectra taken from high flux (upper crosses, black) and low flux
(lower crosses, red) intervals. The bottom panel shows the ratio of high to low
flux spectra, which is consistent with a constant (dashed line).
}
\label{fig:hilow}
\end{figure}

Figure~\ref{fig:hilow} shows a comparison of pn spectra extracted from
intervals of high and low flux. 
These were extracted from only those intervals when the source count
rate was above 10 ct s$^{-1}$ (high) or below 8.8 ct s$^{-1}$
(low). The high and low flux spectra contained $9.6 \times 10^{4}$~ct
and $8.0  \times 10^{4}$~ct, respectively. The ratio of high to low
flux spectra is consistent with a constant ($\chi^{2}=81.6/84~dof$),
again indicating a lack of flux-correlated spectral variability.

\begin{figure}
\centering
\includegraphics[width=6.5 cm, angle=270]{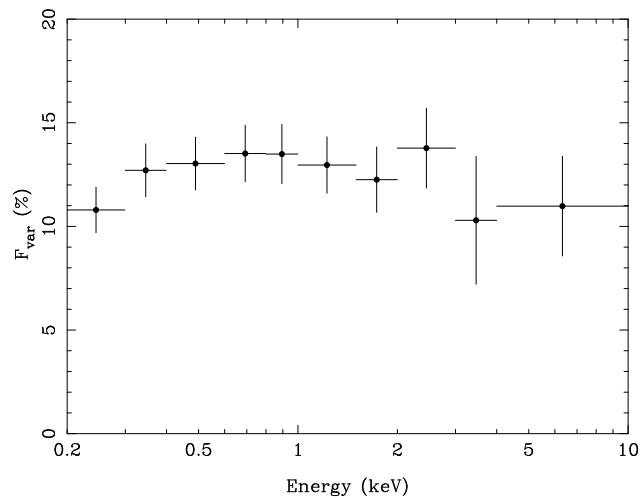}
\caption{
Normalised RMS spectrum based on EPIC pn light curves. The $F_{\rm var}$
values were calculated using 500~s binned data.
}
\label{fig:rms}
\end{figure}

The normalised RMS spectrum was calculated in order to quantify the degree of
variability in the different energy bands, again using the 500~s binned
light curves. The fractional variability amplitude was measured using
the $F_{\rm var}$ statistic (Edelson \et\ 2002) which is equal to the
square root of the normalised excess variance $\sigma_{\rm XS}^2$ (Nandra
\et\ 1997). The errors were
estimated using the formula in Edelson \et\ (2002) and should be
considered as conservative estimates on the true uncertainties.
Figure~\ref{fig:rms} shows the fractional
variability amplitude calculated in ten energy bands, which changes by only 3
per cent
over the \xmm\ band-pass.  This remarkably flat RMS spectrum 
contrasts with the situation in many `normal' spectrum Seyfert 1s,
which tend to show larger fractional variations at lower energies
(e.g. Nandra \et\ 1997). Altering the sizes of the time bins or the
energy ranges used in the light curves made little difference to the
RMS spectrum since most of the variability
power is on timescales $> 500$~s.

\begin{figure}
\centering
\includegraphics[width=6.5 cm, angle=270]{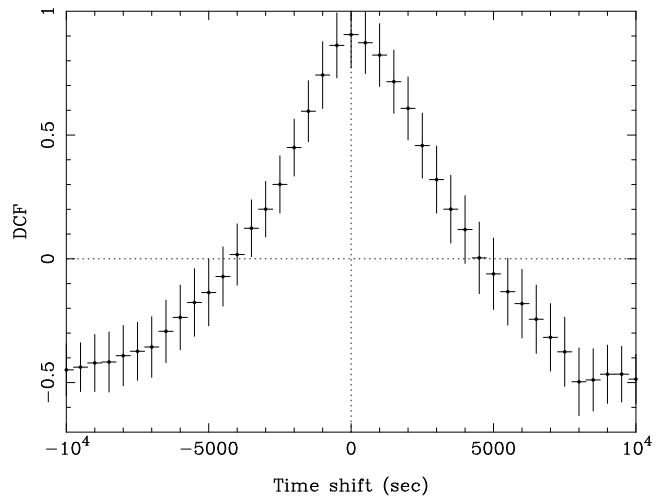}
\caption{
Temporal cross-correlation function for the 0.2--0.3~keV and 0.8--1.0~keV
EPIC pn light curves (using 500~s bins), showing a peak at zero lag. 
}
\label{fig:ccf}
\end{figure}

Having established that the light curves show similar trends and
almost identical amplitudes, the temporal cross-correlation functions
(CCFs) were calculated in order to search for leads or lags between
different energy bands. The Discrete Correlation Function (DCF) of
Edelson \& Krolik (1988) was used as an estimator of the CCF. 
The other nine energy bands used in the RMS spectrum were 
compared to the softest band (0.2--0.3~keV).  Of the
nine CCFs examined eight peaked at zero-lag (e.g., see
figure~\ref{fig:ccf}) and the one exception ($0.2-0.3$~keV against
$1.5-2.0$~keV) peaked at a lag of only one time bin (500~s, with the
harder band lagging the softer band) which we do not consider to be
significant. In all cases the CCF was reasonably symmetric.   These
results indicate there is no evidence for a lag or lead between
the various energy bands to a limit of $\pm 500$~s.

\section{Discussion}
\label{sect:disco}

This paper presents an uninterrupted $\sim 30$~ksec \xmm\ observation
of the ultrasoft, narrow-line Seyfert 1 galaxy \ton. The spectrum
above 2.5~keV is slightly steeper than the norm for Seyfert 1 galaxies
($\Gamma \approx 2.26$ compared to $1.9$) but the iron K$\alpha$ line
emission is only poorly constrained. At lower energies the spectrum
steepens, seen as a soft excess over the hard X-ray power-law with a
featureless, approximately power-law form.  The shape of the soft
excess is similar to that  observed in other objects. For example,
Marshall \et\ (2002) found the high-resolution soft X-ray spectrum of
the ultrasoft Seyfert 1 Mrk~478 (which has a similar X-ray luminosity
to \ton) was a featureless power-law. PKS 0558--504 also showed a
similar soft excess shape (O'Brien \et\ 2001). The \chandra\ HETGS
observation of the variable, narrow-line Seyfert 1 galaxy NGC~4051
again showed the soft excess to be a rapidly variable continuum
component. However, in this lower-luminosity source it showed significant
spectral curvature (Collinge \et\ 2001).

The observed flux of \ton\ showed rapid variability, changing by $\sim
50$ per cent during the observation, but showed only weak spectral
variability. This implies that, to first order at least, the curved
X-ray spectrum is varying as one component. There are indications that
this may be a general result,  that Seyferts with steep X-ray spectra
show very little spectral variability (at least on short
timescales). The \asca\ monitoring of Ark~564 and \ton\ showed flat
RMS spectra (Turner \et\ 2001c; Romano \et\ 2001; Edelson \et\ 2002),
as did the recent \xmm\ observation of 1H~0707--495 (Boller \et\
2002).  This lack of short timescale spectral variability is somewhat
different from the case in ``normal'' Seyfert 1s, which tend to show
stronger variability at lower energies (e.g., Nandra \et\ 1997;
Markowitz \& Edelson, 2001; Vaughan \& Edelson, 2001).

The lack of strong spectral features means that the observed form of
the soft excess is consistent with a range of models. The 
data clearly rule out models based on blends of narrow soft X-ray
lines (e.g., Turner \et\ 1991), but leave a variety of continuum
emission mechanisms as plausible alternatives. Reprocessing of primary
X-rays by an ionised accretion disc can produce a strong, smooth
excess similar to that observed  but has difficulty simultaneously
explaining the hard X-ray spectrum (section~\ref{sect:refl_spec}).
The observed lack of spectral variability is also a challenge for this
model. The highly correlated, simultaneous variability between
3--10~keV flux (dominated by the primary X-rays) and softer X-ray
bands (dominated by the reprocessed emission) requires the
reprocessing to produce a near-perfect ``reverberation'' signature with a
delay of $\Delta t <500$~s (section~\ref{sect:timing}).  Assuming a
black hole mass of  $\gs 10^{7} \Msun$ (from the Eddington limit)
this places the reprocessor within $\ls 10 r_{\rm g}$ of the hard
X-ray source.

As first pointed out by Bechtold \et\ (1987) the soft excess is too
hot to be ``bare'' thermal emission from the accretion disc.  The
spectral form is consistent with emission from multiple blackbody
components, but  the derived temperatures and sizes are inconsistent
with those predicted for an accretion disc (unless \ton\ is highly
super-Eddington).  Allowing for Doppler and gravitational shifts does
not significantly alter this result.

Inverse-Compton scattering of soft photons by thermal electrons
provides a more physically satisfying explanation for the broad soft
excess (sections~\ref{sect:compt_spec} and \ref{sect:compt2_spec}), in
which case the seed photon source is consistent with thermal accretion
disc emission.   The harder power-law emission extending to 10~keV
(and presumably beyond) can also be produced by Comptonisation, either
in another purely thermal plasma or by non-thermal electrons in a
plasma with a  hybrid thermal/non-thermal distribution.  Whether this
is dominated by thermal or non-thermal electrons is impossible to tell
without higher energy data; fits using thermal and hybrid
thermal/non-thermal models are comparable in the \xmm\ band, where the
predicted spectra both resemble power-laws.

A problem with the models discussed above is that they have difficulty
explaining the rapid and energy-independent variability of \ton\
(section~\ref{sect:timing}). These constraints force the hard and
soft X-ray producing regions to be in very close causal contact with
one another. If at high accretion rates the disc is puffed-up due to
radiation pressure, then it is possible that the separation between
the hot corona and the optically thick accretion disc can be
negligibly small. In this scenario, some fraction of the accretion
energy is dissipated in the surface layers of the disc, producing hot
electrons which can Comptonise the soft photons from below.  The
accretion disc model of Hubeny \et\ (2001) demonstrates the formation
of a hot upper layer to the accretion disc due to dissipation near the
surface (see their figure 10), and simulations of magnetized
accretion discs by  Miller \& Stone (2000) suggest that even higher
temperatures may be reached, leading to the formation of a true
corona.  Simulations of emergent disc spectra accounting for
dissipation in the upper atmosphere will be presented elsewhere (Ross
\et\ in prep.), but early indications suggest that the observed
strong, smooth soft excess can be reproduced.

We also note the lack of a significant narrow, neutral iron emission
line at 6.4~keV. The formal (90 per cent) upper limit on the equivalent with
of such a line is $60$~eV, similar to that obtained for 1H 0707--495
by Boller \et\ (2002). Again this contrasts with the situation in more
normal Seyfert 1s, which tend to show a narrow 6.4~keV emission line
from distant, cold material (e.g., Yaqoob, George \& Turner, 2002).
However, whether this is due to a systematic difference between
ultrasoft and normal Seyfert 1s remains to be seen.


\section*{ Acknowledgements }
Based on observations obtained with \xmm, an ESA science mission with
instruments and contributions directly funded by ESA Member States and
the USA (NASA). 
We would like to thank the SOC and SSC teams for making possible the
observations and data analysis.
We thank Gareth Griffiths for useful discussions about the EPIC
calibration and Marek Gierli\'{n}ski for help with the \eqpair\ code.
We thank the referee, Kim Page, for useful comments.
SV acknowledges support from PPARC.
DRB acknowledges
financial support from the Commonwealth Scholarship and Fellowship
Plan and the Natural Sciences and Engineering Research Council of
Canada. WNB acknowledges financial support from NASA LTSA grant NAG5-8107 
and NASA grant NAG5-9939.

\bsp
\label{lastpage}

\end{document}